\documentclass[prd,showpacs,preprintnumbers,twocolumn,amsmath,nofootinbib,amssymb]{revtex4}
\usepackage{graphicx,color,dcolumn,booktabs,bm}
\usepackage{epstopdf}
\usepackage{longtable,lscape}
\usepackage{txfonts}
\usepackage{overpic}
\usepackage{float}
\usepackage{amssymb}
\usepackage{epstopdf}
\usepackage{indentfirst}
\usepackage{feynmf}   
\usepackage{slashed}  
\usepackage{cases}
\usepackage{ulem}
\usepackage{color}
\usepackage{float}
\usepackage{array}

\usepackage{booktabs}
\usepackage{multirow}
\usepackage{tikz}
\usepackage{epstopdf}
\usepackage{epsfig,dsfont,amssymb,amsmath,amsfonts,amsbsy,mathrsfs}
\usepackage{multirow}
\usepackage{graphicx}  
\usepackage{float}  
\usepackage{subfigure}  

\usepackage[colorlinks, citecolor=blue,anchorcolor=red,menucolor=red, linkcolor=red,filecolor=red,runcolor=red,urlcolor=blue,frenchlinks=red]{hyperref}
\makeatletter
\@addtoreset{equation}{section}
\makeatother

\newcommand{\nc}{\newcommand}
\nc{\tj}[1]{\textcolor{red}{Tianjie: #1}}
\allowdisplaybreaks
\begin{document}
\title{The effective $\beta$ value in a Simple Harmonic Oscillator wave function  }
\author{Chao-Hui Wang$^{1,2}$}
\author{Long Tang$^{1,2}$}
\author{Ting-Yan Li$^{1,2}$}
\author{Gong-Ping Zheng$^{1,2}$}
\author{Jing-Fu Hu$^{1,2}$}
\author{Cheng-Qun Pang$^{1,2,3}$\footnote{Corresponding author}}\email{xuehua45@163.com}

\affiliation{$^1$College of Physics and Electronic Information Engineering, Qinghai Normal University, Xining 810000, China\\$^2$Joint Research Center for Physics,
Lanzhou University and Qinghai Normal University,
Xining 810000, China\\$^3$ Lanzhou Center for Theoretical Physics, Key Laboratory of Theoretical Physics of Gansu Province, Lanzhou University, Lanzhou,
Gansu 730000, China  }
\begin{abstract}
When a  Simple Harmonic Oscillator (SHO) wave function is used as an effective wave function, a very important parameter in the SHO wave function is the effective $\beta$ value. We   obtain the analytical expression of $\beta_{eff}$ ($\beta_{effective}$) of the SHO wave function  in coordinate space and momentum space. The expression is applied to the light meson system $(u\bar{u},~u\bar{s})$ to compare the behavior of $\beta_{eff}$.
The results show that $\beta_ {eff,\mathbf{r}}$ in coordinate space and $\beta_ {eff,\mathbf{p}}$ in momentum space are significantly different in the ground state, however, similar in the highly excited states  with Cornell potential.
\end{abstract}
\date{\today}
\maketitle

\section{Introduction}\label{sec1}

Since the quark model \cite{Gell-Mann:1964ewy,zweig1964_3} was proposed, the natures of hadrons have been studied.  The study of the mass and decay of hadron are the most common and important methods for studying the properties of hadron. Different potential models have been proposed for the study of mass spectra, such as the Non-relativistic Cornell potential model \cite{Eichten:1978tg,Eichten:1974af,Lucha:1991vn} and the relativistic Godfrey-Isgur(GI) model \cite{Godfrey:1985xj}. For calculating the strong decay allowed by Okubo-Zweig-Iizuka (OZI) rule, Quark-Pair Creation (QPC) model is a very successful model.

The wave function of the Simple Harmonic Oscillator (SHO) is widely used \cite{Godfrey:1985xj,Guo:2019wpx,Pang:2014laa,Pang:2015eha,Wang:2014sea,Ye:2012gu,Yu:2011ta,He:2013ttg, Wang:2012wa,Liu:2009fe, Sun:2009tg,Sun:2010pg,Li:2009qu,Qin2015Charmed,Isgur:1988gb} for it have similar analytical expressions in coordinate space and momentum space in calculating the energy spectra and decays of hadrons.
\begin{equation}
 \begin{split}
 \begin{cases}
	\Psi_{nLM_L}(\mathbf{r})=R_{nL}(r, \beta)Y_{LM_L}(\Omega_r),\\
	\Psi_{nLM_L}(\mathbf{p})=R_{nL}(p, \beta)Y_{LM_L}(\Omega_p),
\end{cases}
\end{split}
\end{equation}
with
\begin{align} \label{1.2}
R_{nL}(r,\beta)=\beta^{3/2}N_{nL}(\beta r)^{L}e^{\frac{-r^2 \beta^2}{2}} \times L_{n}^{L+1/2}(\beta^2r^2),
\end{align}
\begin{align} \label{1.3}
R_{nL}(p,\beta)=\frac{(-1)^n(-i)^L}{\beta^{3/2}}N_{nL}e^{-\frac{p^2}{2\beta^2}}{(\frac{p}{\beta})}^{L} \times L_{n}^{L+1/2}(\frac{p^2}{ \beta ^2}),
\end{align}
\begin{align} \label{1.4}
N_{nL}=\sqrt{\frac{2n!}{\Gamma(n+L+3/2)}},
\end{align}
where $Y_{LM_L}(\Omega)$ is a spherical harmonic function, $L_{n-1}^{L+1/2}(x)$ is  the associated Laguerre polynomial and $\Gamma(n+L+3/2)$ is the Gamma function \cite{2002Mathematical}.

There is only one parameter  $\beta$ in the SHO wave function. If a wave function is expanded with a set of complete SHO bases, theoretically the solution of the potential model has nothing to do with the value of the parameter $\beta$. However, in actual calculations the number of basis vector we chose  is limited, which makes $\beta$ take an appropriate value.  The selection principle is that the ground state mass calculated under this $\beta$ value is the smallest (for meson $\beta$ value is generally between 0.3-1.5 GeV).

For example, in the decay of highly partial wave in the QPC model, the amplitude has a strong dependence on the shape of the wave function, because of the nodal effect,  i. e. the $\beta$ value has a great impact on the numerical result of the width \cite{Duan:2020tsx}.

There are two options when choosing the $\beta$  value. One is to select a common value (generally 0.4 GeV) and the other is to select the $\beta_{eff}$ value.
In order to gain some properties that cannot be obtained by the  numerical wave function, we prefer to use a  SHO wave function as an effective wave function \cite{Guo:2019wpx,Pang:2014laa,Pang:2015eha,Wang:2014sea,Ye:2012gu,Yu:2011ta,He:2013ttg, Wang:2012wa,Liu:2009fe, Sun:2009tg,Sun:2010pg,Li:2009qu,Isgur:1988gb}. In addition, compared with the vector basis expansion method, the mathematical expression using a SHO wave function as an effective wave function is simpler. When the wave function of a SHO is approximated as the effective wave function of a meson, it is necessary to determine the $\beta_{eff}$ value.

The $\beta_{eff}$ is determined by reproducing the realistic mean square radius  (coordinate space) \cite{Guo:2019wpx,Pang:2014laa,Pang:2015eha,Wang:2014sea,Ye:2012gu,Yu:2011ta,He:2013ttg, Wang:2012wa,Liu:2009fe, Sun:2009tg,Sun:2010pg,Li:2009qu,Isgur:1988gb} and mean square momentum (momentum space) \cite{Qin2015Charmed}, respectively, i. e.
\begin{footnotesize}
\begin{equation}\label{1.5}
\begin{split}
 \begin{cases}
\int |{\psi_{n_r LM}^{SHO}}(r,\beta_{eff,\mathbf{r}})|^2 r^2 sin \theta dr d \theta d \varphi =\int {|\phi(r)}|^2 r^2 dr sin \theta d \theta d \varphi,\\
\int |{\psi_{n_r LM}^{SHO}}(p,\beta_{eff,\mathbf{p}})|^2 p^2  sin \theta dp d \theta d \varphi =\int {|\phi(p)}|^2 p^2 dp sin \theta d \theta d \varphi.
 \end{cases}
\end{split}
\end{equation}
\end{footnotesize}

This paper is organized as follows. Sec. \ref{sec2} is the simple derivation process of the analytical expression of $\beta_{eff}$ in coordinate space and momentum space. In {Sec. \ref{sec3}}, the expressions derived from Sec. \ref{sec2}  are applied to the meson system.
The paper ends with a  conclusion in Sec. \ref{sec4}.

\section{Analytic expression of the $\beta_{eff}$ value}\label{sec2}
In this section, we derive the analytical expression of $\beta_{eff}$ in coordinate space and momentum space.
\subsection{Coordinate space}
As we known, $\beta_{eff}$ in coordinate space is defined
 by the  equation (\ref{1.5}).
\begin{align}\label{2.1}
\int {|\psi_{n_r LM}^{SHO}}(r,\beta_{eff,\mathbf{r}})|^2 r^2 d^3r =\int {|\phi(r,\beta)}|^2 r^2 d^3r ,
\end{align}
where
\begin{equation} \label{2.2}
\phi(r,\beta)=\sum_{i=0}^{N-1}c_{i}u_i(r,\beta).
\end{equation}
The left side of Eq. (\ref{2.1}) $\psi_{n_r LM}^{SHO}(\mathbf{r},\beta_{eff,\mathbf{r}})$  is the effective wave function approximated by a single SHO wave function, on the right slide of   $\phi(r,\beta)$ is the wave function expanded by the harmonic oscillator basis vector $u_i(\mathbf{r},\beta)(\psi^{SHO}(r,\beta))$. The expansion coefficient $c_{i}$ is the eigenvector of the Hamiltonian corresponding to the wave function. The wave function can be written as the product of the radial wave function $R_{nL}(r, \beta)$ and the spherical harmonic function $Y_{LM_L}(\Omega_r)$. Moreover the spherical harmonic function is normalized. Then the formula (\ref{2.1}) can be changed to the following form:
\begin{equation}\label{2.3}
\int r^2 |{R_{n_r~L}^{SHO}}(r,\beta_{eff,\mathbf{r}})|^2 r^2 dr =\int r^2 \sum_{n=0}^{N-1}c_{n}^* c_{n} |{R_{nL}(r,\beta)}|^2 r^2 dr.
\end{equation}
The right side of the formula (\ref{2.3}) can be written as
\begin{equation}\label{2.4}
\begin{split}
\overline{ r^2 }=&\int_{0}^{\infty}\sum_{n,m=0}^{N-1} r^2 c_{n}^* R_{nL}^*(r,\beta) c_{m} R_{mL}(r,\beta)r^2~dr\\
=&\int_{0}^{\infty}\sum_{n,m=0}^{N-1} r^2 c_{n}^* \beta^{3/2}N_{nL}(\beta r)^{L}e^{\frac{-r^2 \beta^2}{2}} \times L_{n}^{L+1/2}(\beta^2r^2) \\
&\times c_{m}\beta^{3/2}N_{mL}(\beta r)^{L}e^{\frac{-r^2 \beta^2}{2}} \times L_{m}^{L+1/2}(\beta^2r^2) r^2~dr.\\
\end{split}
\end{equation}
$\beta{r}$ can be substituted with $z$,  and  $\overline{ r^2 }$ expressed as:
\begin{equation}\label{2.5}
\begin{split}
\overline{ r^2 }=&\sum_{n,m=0}^{N-1}c_{n}^* c_{m} \frac{1}{2\beta^2}  N_{nL} N_{mL}  \int_{0}^{\infty} e^{-z^2} \times L_{n}^{L+\frac{1}{2}}(z^2)\\
& \times L_{m}^{L+1/2}(z^2) {(z^2)}^{L+\frac{3}{2}}  dz^2.
\end{split}
\end{equation}
We replace $z^2$ with $x$,
\begin{equation}\label{2.6}
\begin{split}
\overline{ r^2 }=&\sum_{n,m=0}^{N-1}c_{n}^* c_{m} \frac{1}{2\beta^2}  N_{nL} N_{mL}  \int_{0}^{\infty} e^{-x} x^{L+\frac{3}{2}} \\
& \times L_{n}^{L+1/2}(x) \times L_{m}^{L+1/2}(x) dx.
\end{split}
\end{equation}
Because Laguerre polynomials have the property \cite{2002Mathematical}: $L_{n}^u(z)=L_{n}^{u+1}(z)-L_{n-1}^{u+1}(z)$, (\ref{2.6}) can take the form:
\begin{equation}\label{2.7}
\begin{split}
\overline{ r^2 }=&\sum_{n,m=0}^{N-1}c_{n}^* c_{m} \frac{1}{2\beta^2}  N_{nL} N_{mL}  \int_{0}^{\infty} e^{-x} x^{L+\frac{3}{2}}\times [(L_{n}^{L+3/2}(x)-\\
&L_{n-1}^{L+3/2}(x))\times (L_{m}^{L+3/2}(x)-L_{m-1}^{L+3/2}(x))] dx.
\end{split}
\end{equation}
The properties  of Laguerre function and $\Gamma$ function  used in the calculation process of Eq. (\ref{2.7}) are listed below \cite{2002Mathematical} :
\begin{equation}\label{2.8}
\int_{0}^{\infty} e^{-x} x^{\alpha} \times  L_{n}^{\alpha}(x)\times L_{m}^{\alpha}(x) dx=\delta_{nm}\frac{\Gamma(n+\alpha+1)}{n!},
\end{equation}
\begin{equation}\label{2.9}
\Gamma(z+1)=z\Gamma(z).
\end{equation}
When calculating the equation
\begin{equation}\label{2.10}
\begin{split}
\overline{ r^2 }=&\frac{1}{\beta^2}\sum_{n,m=0}^{N-1}c_{n}^* c_{m}[(n+L+3/2)\delta_{n,m}-\sqrt{m(n+L+\frac{3}{2})}\delta_{n,m-1}\\
&-\sqrt{n(m+L+3/2)}
\delta_{n-1,m}+n\delta_{n-1,m-1}],
\end{split}
\end{equation}
we use the property of $\delta$  function  to convert the double sum of the above equation into a single sum form \cite{2002Mathematical}
\begin{equation}\label{2.11}
\begin{split}
\overline{ r^2 }=&\frac{1}{\beta^2}[\sum_{n=0}^{N-1}c_{n}^* c_{n}(2n+L+3/2)\\
&-2\sum_{n=0}^{N-2}c_{n}^*c_{n+1}\sqrt{(n+1)(n+L+\frac{3}{2})}].
\end{split}
\end{equation}
According to the relation (\ref{2.3}) of~~$\beta_ {eff,\mathbf{r}}$, the analytic expression of $\beta_ {eff,\mathbf{r}}$ in coordinate space is obtained
\begin{footnotesize}
\begin{equation}\label{2.12}
\begin{split}
&\beta^2_{eff,\mathbf{r}}=\beta^2(2n_r+L+\frac{3}{2}) \\
&\times \frac{1}{\sum_{n=0}^{N-1}c_{n}^* c_{n}(2n+L+\frac{3}{2})-2\sum_{n=0}^{N-2}c_{n}^*c_{n+1}\sqrt{(n+1)(n+L+\frac{3}{2})}}.
\end{split}
\end{equation}
\end{footnotesize}
\subsection{Momentum space}
\par
Likewise, the $\beta_ {eff}$ in momentum space is produced by the definition:
\begin{footnotesize}
\begin{equation}\label{2.13}
\int p^2 |{R_{n_rL}^{SHO}}(p,\beta_{eff,\mathbf{p}})|^2 p^2 dp =\int p^2 \sum_{n=0}^{N-1}c_{n}^* c_{n} |{R_{nL}(p,\beta)}|^2 p^2 dp.
\end{equation}
\end{footnotesize}
\begin{equation}\label{2.14}
\begin{split}
\overline{ p^2 }=&\int_{0}^{\infty}\sum_{n,m=0}^{N-1} p^2 c_{n}^* R_{nL}^*(p,\beta)  c_{m} R_{mL}(p,\beta)p^2~dp\\
=&\int_{0}^{\infty}\sum_{n,m=0}^{N-1} p^2 c_{n}^* \frac{(-1)^n(i)^L}{\beta^{3/2}}N_{nL}e^{-\frac{p^2}{2\beta^2}}{(\frac{p}{\beta})}^{L} \times L_{n}^{L+1/2}(\frac{p^2}{ \beta ^2})  \\
&\times c_{m} \frac{(-1)^m(-i)^L}{\beta^{3/2}}N_{mL}e^{-\frac{p^2}{2\beta^2}}{(\frac{p}{\beta})}^{L} \times L_{m}^{L+1/2}(\frac{p^2}{ \beta ^2})p^2 dp\\
=&\sum_{n,m=0}^{N-1} c_{n}^*  c_{m} N_{nL}N_{mL} (-1)^{(n+m)}\int_{0}^{\infty}{(\frac{p}{\beta})}^{2L}\frac{p^2}{\beta^2}\frac{p^2}{\beta}
e^{-\frac{p^2}{\beta^2}}\\
&\times L_{n}^{L+1/2}(\frac{p^2}{ \beta ^2})\times L_{m}^{L+1/2}(\frac{p^2}{ \beta ^2}) dp,\\
\end{split}
\end{equation}
when we assume $z=\frac{p}{\beta}$, $\overline{p^2}$  can be written as:
\begin{equation}\label{2.15}
\begin{split}
\overline{ p^2 }
=&\sum_{n,m=0}^{N-1} c_{n}^*  c_{m} N_{nL}N_{mL} (-1)^{(n+m)}\frac{\beta^2}{2}\\
&\times \int_{0}^{\infty}e^{-z^2}{(z^2)}^{L+\frac{3}{2}}\times L_{n}^{L+1/2}(z^2)\times L_{m}^{L+1/2}(z^2) dz^2.\\
\end{split}
\end{equation}
To displace $z^2$ with $x$,
\begin{equation}\label{2.16}
\begin{split}
\overline{ p^2 }
=&\sum_{n,m=0}^{N-1} c_{n}^*  c_{m} N_{nL}N_{mL} (-1)^{(n+m)}\beta^2 \\
&\times\int_{0}^{\infty}e^{-x} x^{L+\frac{3}{2}}
\times L_{n}^{L+1/2}(x)\times L_{m}^{L+1/2}(x) dx.\\
\end{split}
\end{equation}
The equation (\ref{2.16})  is  similar to (\ref{2.6}).  Thus we can obtain the analytical expression  of $\overline{p^2}$ in momentum space. For the following eqution:
\begin{equation}\label{2.17}
\begin{split}
\overline{ p^2 }=&\beta^2\sum_{n,m=0}^{N-1}(-1)^{(n+m)} c_{n}^*  c_{m} [(n+L+3/2)\delta_{n,m}\\
&-\sqrt{m(n+L+\frac{3}{2})}\delta_{n,m-1}-\sqrt{n(m+L+3/2)}\delta_{n-1,m}\\
&+n\delta_{n-1,m-1}],
\end{split}
\end{equation}
 we use the previous schedule to get the formula:
\begin{equation}\label{2.18}
\begin{split}
\overline{ p^2 }= &\beta^2[\sum_{n=0}^{N-1}c_{n}^* c_{n}(2n+L+3/2)\\
&+2\sum_{n=0}^{N-2}c_{n}^*c_{n+1}\sqrt{(n+1)(n+L+\frac{3}{2})}].
\end{split}
\end{equation}
Then, according to the relation (\ref{2.13}) of $\beta_ {eff}$, the analytic expression of $\beta_{eff,\mathbf{p}}$ in momentum space is obtained
\begin{equation}\label{2.19}
\begin{split}
\beta_{eff,\mathbf{p}}^2=&\frac{\beta^2}{(2n_r+L+3/2)}\times
[\sum_{n=0}^{N-1}c_{n}^* c_{n}(2n+L+3/2)\\&+2\sum_{n=0}^{N-2}c_{n}^*c_{n+1}\sqrt{(n+1)(n+L+\frac{3}{2})}].
\end{split}
\end{equation}

We obtain the analytic expression of $\beta_{eff}$ in coordinate space and momentum space respectively.
Here, we make
\begin{equation}\label{3.3}
\begin{split}
\begin{cases}
k_1=\sum_{n=0}^{N-1}c_{n}^* c_{n}(2n+L+\frac{3}{2}), \\
k_2=2\sum_{n=0}^{N-2}c_{n}^*c_{n+1}\sqrt{(n+1)(n+L+\frac{3}{2})}.
\end{cases}
\end{split}
\end{equation}
Then equations (\ref{2.12}) and (\ref{2.19}) can be written as
\begin{equation}\label{3.4}
\begin{split}
\begin{cases}
&\beta^2_{eff,\mathbf{r}}=\frac{\beta^2(2n_r+L+\frac{3}{2})}{k_1-k_2},\\
&\beta_{eff,\mathbf{p}}^2=\frac{\beta^2(k_1+k_2)}{2n_r+L+\frac{3}{2}}.
\end{cases}
\end{split}
\end{equation}

\subsection{ $\beta_{eff,\mathbf{r}}$ and $\beta_{eff,\mathbf{p}}$ analytical verification}\label{sec2.3}
If we take the wave function of hydrogen atom as the strict wave function, it can be used to test whether equation (\ref{3.4}) is correct. The wave function of hydrogen atom is composed of its radial wave function and angular wave function. The normalized radial wave function of hydrogen atom is
\begin{footnotesize}
\begin{equation}
R_{n l}(r)=\sqrt{\left(\frac{2}{a n}\right)^{3} \frac{(n-l-1) !}{2 n(n+l) !}}\left(\frac{2 r}{a n}\right)^{l} \exp \left(-\frac{r}{a n}\right) \mathrm{L}_{n-l-1}^{2 l+1}\left(\frac{2 r}{a n}\right), \label{b1.1}
\end{equation}
\end{footnotesize}
here, $a\equiv\frac{\hbar}{m e^2}=\frac{1}{m\alpha}$ is the Bohr radius of the hydrogen atom  ($m$ is the electron mass, 0.511 MeV,  $e=\sqrt{\alpha}$ is the charge of an electron, and $\alpha=1/137$ is the Fine Structure Constant.).  We know that the energy level only depends on a special combination of radial quantum number $n_r$ and angular momentum quantum number $L$, that is, it only depends on the main quantum number $n = n_r + L + 1$.

Then Using the definition of the mean value of mechanical quantity operator, we obtain the analytical formula of the $\overline{r^2}$ mean value of hydrogen atomic wave function.
\begin{equation}
\overline{r^2}=\frac{1}{2}a^2n^2[6 + 2 L^2 + 10 n_r + 5 n^2_r + L (7 + 10 n_r)].
\end{equation}
Due to the complexity of calculating the average value of momentum square of hydrogen atomic wave function, Feynman-Hellmann theorem is used to calculate $\overline{p^2}$ \cite{Feynman:1939zza,None1935Acta,Quigg:1979vr}. In the wave function of hydrogen atom, the potential energy is Coulomb potential, according to Feynman-Hellmann  theorem, there are
\begin{equation}
\frac{\overline{p^2}}{2m}=\overline{T}=\frac{-1}{2-1}E=-(-\frac{e^2}{2an^2}).
\end{equation}
And $a=\frac{\hbar}{me^2}$, so we obtain $\overline{p^2}$
\begin{equation}
\overline{p^2}=\frac{1}{a^2n^2}.
\end{equation}
By analogy (\ref{1.5}), we can obtain the $\beta_{eff}$ analytical formula when the hydrogen atom wave function is used as the exact wave function.
\begin{equation}\label{add1.2}
\begin{split}
 \begin{cases}
 \frac{(2n_r+L+\frac{3}{2})}{\beta_{eff,r}^2}=\frac{1}{2}a^2n^2 [6 + 2 L^2 + 10 n_r + 5 n^2_r + L (7 + 10 n_r)], \\
\beta_{eff,p}^2\cdot(2n_r+L+3/2)=\frac{1}{a^2n^2}.
\end{cases}
\end{split}
\end{equation}

Through equations (\ref{add1.2}), we can easily get the relationship of  $\beta_{eff,\bf{r}}$ and $\beta_{eff,\bf{p}}$

\begin{equation}
\begin{split}
 \begin{cases}
\beta_{eff,r}^2=\frac{2(2n_r+L+\frac{3}{2}) }{ a^2n^2 [6 + 2 L^2 + 10 n_r + 5 n_r^2 + L (7 + 10 n_r)]}, \\
\beta_{eff,p}^2=\frac{1}{a^2n^2(2n_r+L+3/2)}. \label{add1.3}
\end{cases}
\end{split}
\end{equation}

\begin{equation}\label{add1.4}
\frac{\beta_{eff,r}}{\beta_{eff,p}}=\frac{\sqrt{2}(2n_r+L+\frac{3}{2})}{\sqrt{6 + 2 L^2 + 10 n_r + 5 n_r^2 + L (7 + 10 n_r)}}.
\end{equation}

From equation (\ref{add1.4}), we can get the following two conclusions when the hydrogen atom wave function is used as the exact wave function to calculate $\beta_{eff,r}/\beta_{eff,p}$:

\begin{itemize}
\item When $L$ is finite and $n_r$ tends to infinity, $\beta_{eff,r}/\beta_{eff,p}=2\sqrt{\frac{2}{5}}\approx 1.2649$.
\item When $n_r$ is finite and $L$ tends to infinity, $\beta_{eff,r}/\beta_{eff,p}=1$.
\end{itemize}

\begin{figure}[htp]
\begin{center}
	\centerline{\includegraphics[width=235pt]{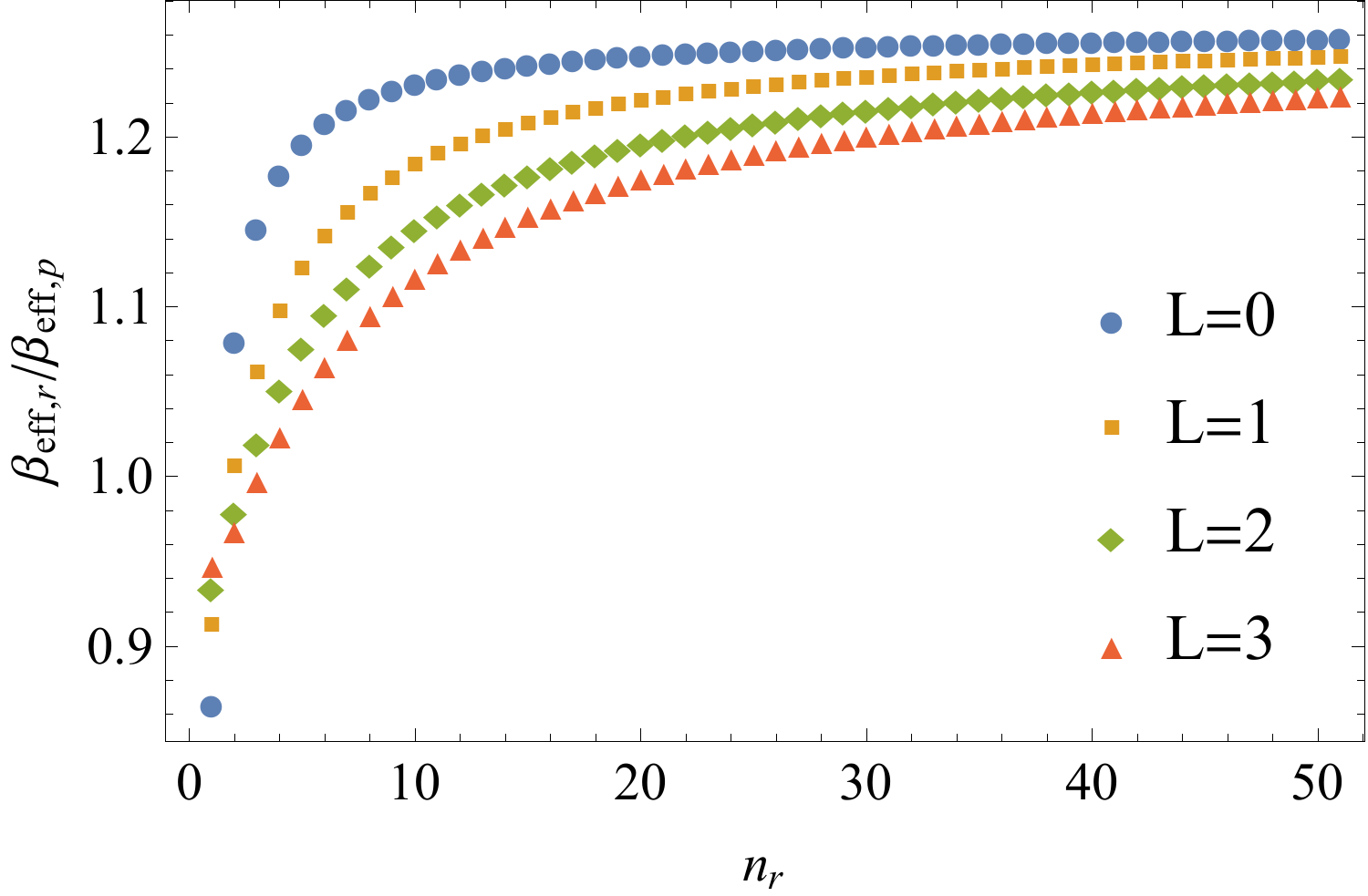}}
\caption{The curve of $\beta_{eff,r}/\beta_{eff,p}$ when $L =(0,3), n_r=(0,50)$.}
\label{nr(0-50)}
\end{center}
\end{figure}

\begin{figure}[htp]
\begin{center}
	\centerline{\includegraphics[width=235pt]{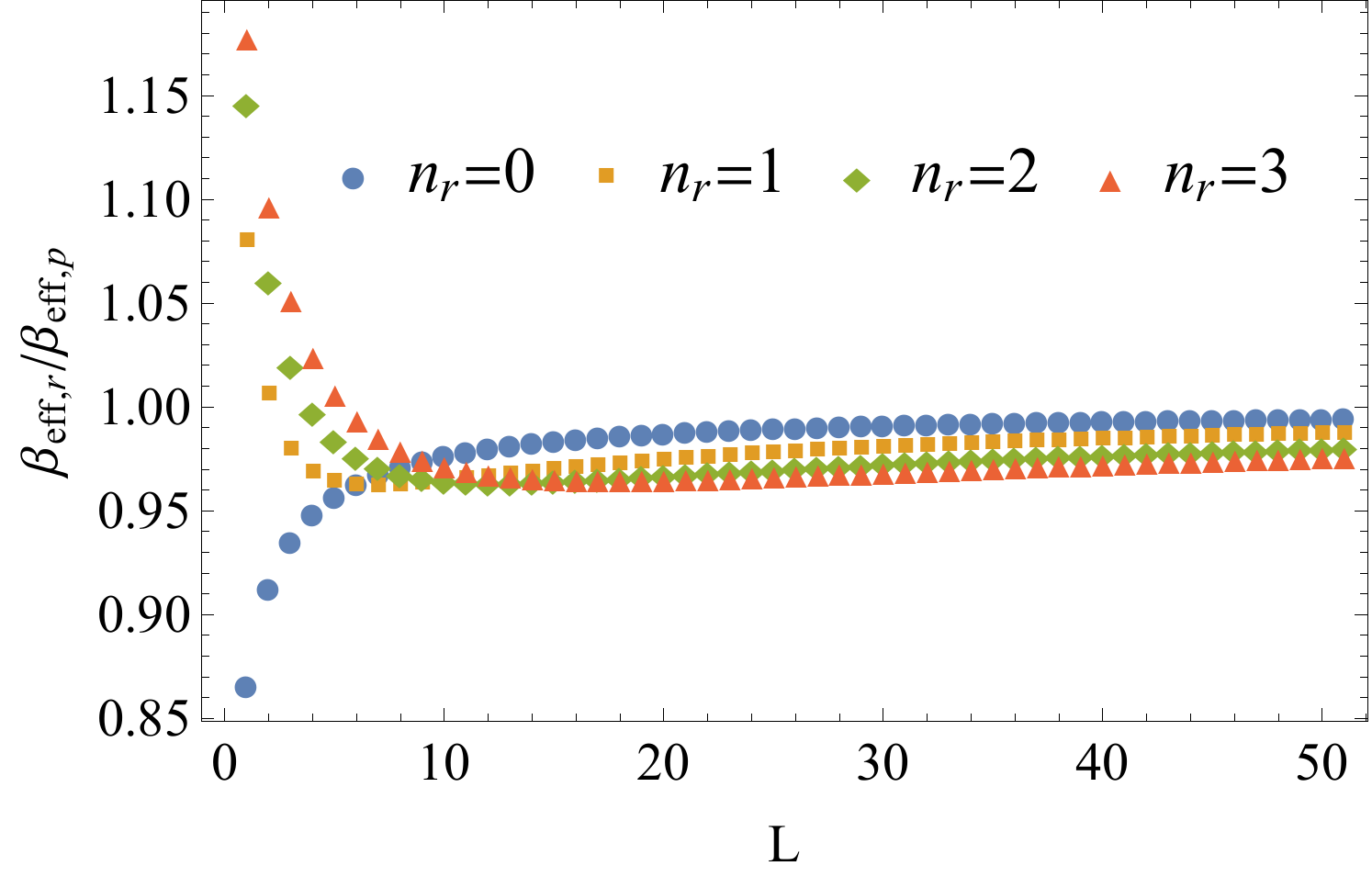}}
\caption{The curve of $\beta_{eff,r}/\beta_{eff,p}$ when $n_r =(0,3), L=(0,50)$.}
\label{L(0-50)}
\end{center}
\end{figure}

We draw the curve of $\beta_{eff,r}/\beta_{eff,p}$ when $n_r =(0,3)$, $L=(0,50)$   and $L =(0,3)$, $n_r=(0,50)$, respectively, as shown in Fig. \ref{nr(0-50)} and Fig. \ref{L(0-50)}. In Fig. \ref{nr(0-50)} and Fig. \ref{L(0-50)}, when we take the larger value of $n_r(L)$, the value of $\beta_{eff,r}/\beta_{eff,p}$ is closer to $1.2649 (1)$.
In addition, we bring these formulas into the Hamiltonian equation satisfied by the wave function of hydrogen atom to obtain $\beta_{eff,\bf{r}}$, $\beta_{eff,\bf{p}}$ and $\beta_{eff,\bf{r}}/\beta_{eff,\bf{p}}$, the result is list in Tab. \ref{hydrogen atom}. It can be seen from Fig. \ref{nr(0-50)}, Fig. \ref{L(0-50)} and Tab. \ref{hydrogen atom} that in most cases, when $n_r< L$, $\beta_{eff,r}/\beta_{eff,p}< 1$, and when $n_r >L$, $\beta_{eff,r}/\beta_{eff,p}> 1$. In  Tab. \ref{hydrogen atom}, we use the result of the exact solution of the  $\beta_{eff}$ value Eq. (\ref{add1.3}) under the hydrogen atom wave function to verify the result of the  $\beta_{eff}$ value of the  Eq.  (\ref{3.4}) when the harmonic oscillator basis is expanded. The data in the Tab. \ref{hydrogen atom} show that the  Eq. (\ref{3.4}) is correct. (It should be noted that in the calculation process of the basis vector expansion in this paper, the size of the matrix taken is $16\times 16$. If the number of matrix basis vectors is more, the data will be more accurate.)

\begin{table*}[htp]
\centering
 \footnotesize
	\caption{ $\beta_{eff,\mathbf{r}}, \beta_{eff,\mathbf{p}}$ and energy spectrum in the wave function of hydrogen atom. (unit: keV)  }
\label{hydrogen atom}
\[\begin{array}{ccc|cccc|cccc}
 \toprule[1pt]  \toprule[1pt]
\multicolumn{3}{c|}{}  &\multicolumn{4}{|c|}{Exact~~solution }  &\multicolumn{4}{|c}{Numerical~~solution(SHO-expednded)}  \\
\hline
n_r & L & state~(nl)  & \beta_{eff,\mathbf{r}} & \beta_{eff,\mathbf{p}} & \beta_{eff,\mathbf{r}}/ \beta_{eff,\mathbf{p}} & energy\times 10^{-3} & \beta_{eff,\mathbf{r}} & \beta_{eff,\mathbf{p}} & \beta_{eff,\mathbf{r}}/ \beta_{eff,\mathbf{p}} & energy\times 10^{-3} \\
\hline
0  &  0  & 1S & 2.6375 & 3.0455 & 0.8660 & -13.6129 & 2.6630 & 3.0399 & 0.8760 & -13.5639\\
0  &  1  & 2P & 1.0767 & 1.1795 & 0.9129 & -3.4032  & 1.0790 & 1.1790 & 0.9151 & -3.4013 \\
0  &  2  & 3D & 0.6217 & 0.6646 & 0.9354 & -1.5125  & 0.6220 & 0.6645 & 0.9361 & -1.5123 \\
0  &  3  & 4F & 0.4170 & 0.4396 & 0.9487 & -0.8508  & 0.4172 & 0.4396 & 0.9490 & -0.8508 \\
0  &  4  & 5G & 0.3045 & 0.3181 & 0.9574 & -0.5445  & 0.3046 & 0.3181 & 0.9576 & -0.5445 \\
1  &  0  & 2S & 1.0767 & 0.9969 & 1.0801 & -3.4032  & 1.0903 & 0.9886 & 1.1030 & -3.4032 \\
1  &  1  & 3P & 0.5898 & 0.5861 & 1.0062 & -1.5125  & 0.5932 & 0.5847 & 1.0145 & -1.5125 \\
1  &  2  & 4D & 0.3896 & 0.3976 & 0.9800 & -0.8508  & 0.3908 & 0.3973 & 0.9836 & -0.8508 \\
1  &  3  & 5F & 0.2835 & 0.2926 & 0.9690 & -0.5445  & 0.2839 & 0.2925 & 0.9708 & -0.5445 \\
2  &  0  & 3S & 0.6080 & 0.5301 & 1.1468 & -1.5125  & 0.6089 & 0.5158 & 1.1806 & -1.5125 \\
2  &  1  & 4P & 0.3882 & 0.3657 & 1.0614 & -0.8508  & 0.3924 & 0.3624 & 1.0827 & -0.8508 \\
2  &  2  & 5D & 0.2780 & 0.2724 & 1.0206 & -0.5445  & 0.2806 & 0.2715 & 1.0334 & -0.5445 \\
\midrule[1pt] \midrule[1pt]
\end{array}\]\label{tab5}
\end{table*}

\section{Discussion of~$\beta_ {eff,\mathbf{r}}$~and~$\beta_ {eff,\mathbf{p}}$ }\label{sec3}

\subsection{ $\beta_{eff,\mathbf{r}}$ and $\beta_{eff,\mathbf{p}}$ behavior with Cornell potential}

In order to make a comparison with the numerical results of Close \cite{Close:2005se}, the potential and parameters used in the calculation are consistent with those in reference \cite{Close:2005se}:
\begin{equation}\label{3.5}
V(r)=\frac{4}{3}C_{q\bar{q}}-\frac{4}{3}\frac{\alpha_s}{r}+br+\frac{32\pi \alpha_s }{9 m_1 m_2}\tilde
{\delta_{\sigma}}(r)\vec{S}_1\cdot \vec{S}_2,
\end{equation}
where  $\tilde{\delta}_{\sigma}(r)=(\frac{\sigma}{\sqrt{\pi}})^3e^{-\sigma^2r^2}$.
The parameters used for the calculation are listed as follows \cite{Close:2005se}: $C_{us}=-0.500$~GeV, $C_{uu}=-0.330$~GeV, b= 0.162$~\rm  GeV^2$, $\alpha$ = 0.594, $\sigma$ = 0.897~GeV, $m_u$ = 0.33~GeV, $m_c$ = 1.6~GeV . The following calculations are listed as an example of a system consisting of $u\bar{u}$ quark and $u\bar{s}$ quark.

The behavior of $\beta_{eff}$  is discussed below. Our analytical results are compared with those in Ref. \cite{Close:2005se}.

\subsection{A brief introduction of the QPC model}
The QPC model was first proposed by Micu \cite{Micu:1968mk}. After decades of further development \cite{LeYaouanc:1972vsx,LeYaouanc:1973ldf,LeYaouanc:1974cvx,LeYaouanc:1977gm,LeYaouanc:1977fsz,vanBeveren:1982qb,Bonnaz:2001aj}, the model has been widely used to calculate the strong two-body decay allowed by Okubo - Zweig - Iizuka (OZI).
In the QPC model, the transition matrix of the decay process $A\to B+C$  is defined by
\begin{equation}\label{3.6}
\langle BC|\mathcal{T}|A \rangle = \delta ^3(\mathbf{P}_B+\mathbf{P}_C)\mathcal{M}^{{M}_{J_{A}}M_{J_{B}}M_{J_{C}}},
\end{equation}
where $\mathcal{M}^{{M}_{J_{A}}M_{J_{B}}M_{J_{C}}}$ is the amplitude of $A\to B+C$. $\mathcal{T}$ is the transition operator, which can describe the quark-antiquark pair generated from the vacuum. It can be expressed as
\begin{align}\label{3.7}
\mathcal{T}& = -3\gamma \sum_{m}\langle 1m;1~-m|00\rangle\int d \mathbf{p}_3d\mathbf{p}_4\delta ^3 (\mathbf{p}_3+\mathbf{p}_4) \nonumber \\
 & ~\times \mathcal{Y}_{1m}\left(\frac{\textbf{p}_3-\mathbf{p}_4}{2}\right)\chi _{1,-m}^{34}\phi _{0}^{34}
\left(\omega_{0}^{34}\right)_{ij}b_{3i}^{\dag}(\mathbf{p}_3)d_{4j}^{\dag}(\mathbf{p}_4){,}
\end{align}
in the formula (\ref{3.7}), the subscripts 3 and 4  represent quark and antiquark, respectively. $\chi$, $\phi$ and $\omega$ represent spin, taste and color wave functions, respectively. $\gamma$ is a dimensionless constant, which describes the generation rate of positive and negative quark pairs in vacuum, and it is generally determined by fitting experimental data.  $\mathcal{Y}_{\ell m}(\mathbf{p})={|\mathbf{p}|^{\ell}}Y_{\ell
m}(\mathbf{p})$ is the solid harmonic. According to the Jacobi-Wick formula, the helicity amplitude is converted into the partial wave amplitude, and the amplitude is expressed as
\begin{align}\label{3.8}\begin{split}
\mathcal{M}^{J L}(\mathbf{P})=& \frac{\sqrt{4 \pi(2 L+1)}}{2 J_{A}+1} \sum_{M_{J_{B}} M_{J_{C}}}\left\langle L 0 ; J M_{J_{A}} \mid J_{A} M_{J_{A}}\right\rangle \\
& \times\left\langle J_{B} M_{J_{B}} ; J_{C} M_{J_{C}} \mid J_{A} M_{J_{A}}\right\rangle \mathcal{M}^{M_{J_{A}} M_{J_{B}} M_{J_{C}}}.
\end{split}\end{align}
Therefore, the decay width of $A \rightarrow B C$ is read as
\begin{align}\label{3.9}
\Gamma=\frac{\pi}{4} \frac{|P|}{m_{A}^{2}} \sum_{J, L}\left|M^{J L}(\boldsymbol{P})\right|^{2},
\end{align}
where $\mathrm{m}_{A}$ is the mass of the initial state A-meson.

In addition, the meson wave function is defined as mock state, i. e.
\begin{align}\label{3.10}\begin{split}
\left|A\left(n^{2 S+1} L_{J M_{J}}\right)\left(\mathbf{p}_{A}\right)\right\rangle=& \sqrt{2 E} \sum_{M_{S}, M_{L}}\left\langle L M_{L} S M_{S} \mid J M_{J}\right\rangle \chi_{S M_{S}}^{A} \\
& \times \phi^{A} \omega^{A} \int d \mathbf{p}_{1} d \mathbf{p}_{2} \delta^{3}\left(\mathbf{p}_{A}-\mathbf{p}_{1}-\mathbf{p}_{2}\right) \\
& \times \Psi_{n L M_{L}}^{A}\left(\mathbf{p}_{1}, \mathbf{p}_{2}\right)\left|q_{1}\left(\mathbf{p}_{1}\right) \bar{q}_{2}\left(\mathbf{p}_{2}\right)\right\rangle,
\end{split}
\end{align}
and here the spatial wave function $\psi_{n L M_{L}}(\mathbf{p})$ of the meson is the SHO wave function in momentum space.

The QPC model is calculated in momentum space, the $\beta_{eff}$ in wave function here should also be more reasonable from momentum space. However the $\beta_{eff}$ of coordinate space is widely used in previous calculations \cite{Guo:2019wpx,Pang:2014laa,Pang:2015eha,Wang:2014sea,Ye:2012gu,Yu:2011ta,He:2013ttg, Wang:2012wa,Liu:2009fe, Sun:2009tg,Sun:2010pg,Li:2009qu,Isgur:1988gb}. It is necessary to  investigate the difference of results of  the QPC model with $\beta_{eff,\mathbf{r}}$ and $\beta_{eff,\mathbf{p}}$.

\subsection{The comparison between  $\beta_{eff,\textbf{r}}$ and $\beta_{eff,\textbf{p}}$ with Close potential}
\vspace{20pt}
\hspace{60pt}
{{\it{case 1: $\beta$ =\text{0.4 GeV}}}}
\vspace{10pt}

\par When $\beta$ in the exact wave function takes the common value 0.4 GeV as the input, the corresponding $\beta_{eff}$ values are obtained according to Eq. (\ref{3.4}) as shown in Tab. \ref{tab1} and Tab. \ref{tab2}.

\renewcommand{\arraystretch}{1.2}
\begin{table}[htbp]\footnotesize
	\centering
\caption{The $\beta_{eff}$ value of $u\bar{u}$ quark system when $\beta$ takes the common value of 0.4 GeV (unit: GeV).}
\label{uu,0.4}
\[\begin{array}{cccc|cccc}
\toprule[1pt] \toprule[1pt]
state & \beta_{eff,\mathbf{r}} &  \beta_{eff,\mathbf{r}} $ \cite{Close:2005se}$  & \beta_{eff,\mathbf{p}} & state  & \beta_{eff,\mathbf{r}} &  \beta_{eff,\mathbf{r}}$ \cite{Close:2005se}$ &  \beta_{eff,\mathbf{p}}  \\
\hline
0^1S_0       & 0.4760      & 0.47 & 0.5405 & 0^3S_1       & 0.2837      & 0.28 & 0.2850 \\
1^1S_0       & 0.2801      & 0.28 & 0.2710 & 1^3S_1       & 0.2412      & 0.24 & 0.2391 \\
2^1S_0       & 0.2411      & 0.24 & 0.2342 & 2^3S_1       & 0.2229      & 0.23 & 0.2212 \\
3^1S_0       & 0.2269      &      & 0.2228 & 3^3S_1      & 0.2206      &      & 0.2222 \\
0^1P_1       & 0.2715      & 0.27 & 0.2761 & 0^3P_J       & 0.2630      & 0.26 & 0.2644 \\
1^1P_1       & 0.2387      &      & 0.2414 & 1^3P_J      & 0.2326      &      & 0.2320 \\
2^1P_1       & 0.2239      &      & 0.2259 & 2^3P_J       & 0.2195      &      & 0.2190 \\
3^1P_1       & 0.2278      &      & 0.2326 & 3^3P_J       & 0.2258      &      & 0.2296 \\
0^1D_2       & 0.2469      & 0.25 & 0.2479 & 0^3D_J       & 0.2464      & 0.25 & 0.2474 \\
1^1D_2       & 0.2254      &      & 0.2257 & 1^3D_J       & 0.2249      &      & 0.2250 \\
2^1D_2       & 0.2188      &      & 0.2202 & 2^3D_J       & 0.2183      &      & 0.2195 \\
3^1D_2       & 0.2339      &      & 0.2390 & 3^3D_J       & 0.2336      &      & 0.2386 \\
\bottomrule[1pt] \bottomrule[1pt]
\end{array}\]\label{tab1}
\end{table}

\renewcommand{\arraystretch}{1.2}
\begin{table}[htbp]\footnotesize
\centering
\caption{The $\beta_{eff}$ value of $u\bar{s}$ quark system when $\beta$ takes the common value of 0.4 GeV (unit: GeV).}
\label{uc,0.4}
\[\begin{array}{cccc|cccc}
\toprule[1pt] \toprule[1pt]
state  & \beta_{eff,\mathbf{r}}  &  \beta_{eff,\mathbf{r}}$ \cite{Close:2005se}$ &  \beta_{eff,\mathbf{p}} & state  & \beta_{eff,\mathbf{r}} &  \beta_{eff,\mathbf{r}}$ \cite{Close:2005se}$ &  \beta_{eff,\mathbf{p}} \\
\hline
0^1S_0       & 0.4535 & 0.46  & 0.5024 & 0^3S_1       & 0.3155      & 0.32 & 0.3179 \\
1^1S_0       & 0.2942 & 0.29  & 0.2896 & 1^3S_1       & 0.2638      & 0.26 & 0.2616 \\
2^1S_0       & 0.2563 & 0.25  & 0.2513 & 2^3S_1       & 0.2417      & 0.24 & 0.2396 \\
3^1S_0       & 0.2387 &       & 0.2347 & 3^3S_1       & 0.2314      &      & 0.2304 \\
0^1P_1       & 0.2939 & 0.29  & 0.2984 & 0^3P_J       & 0.2859      & 0.29 & 0.2876 \\
1^1P_1       & 0.2574 &       & 0.2595 & 1^3P_J       & 0.2518      &      & 0.2512 \\
2^1P_1       & 0.2392 &       & 0.2401 & 2^3P_J       & 0.2350      &      & 0.2339 \\
3^1P_1       & 0.2335 &       & 0.2354 & 3^3P_J       & 0.2310      &      & 0.2318  \\
0^1D_2       & 0.2673 & 0.27  & 0.2685 & 0^3D_J       & 0.2668      & 0.27 & 0.2679 \\
1^1D_2       & 0.2432 &       & 0.2435 & 1^3D_J       & 0.2426      &      & 0.2427 \\
2^1D_2       & 0.2310 &       & 0.2312 & 2^3D_J       & 0.2304      &      & 0.2304 \\
3^1D_2       & 0.2346 &       & 0.2376 & 3^3D_J       & 0.2342      &      & 0.2370 \\
\bottomrule[1pt] \bottomrule[1pt]
\end{array}\]\label{tab2}
\end{table}

From Table \ref{tab1} and Table \ref{tab2}, it can be seen that $\beta_{eff,\mathbf{r}}$ and $\beta_{eff,\mathbf{p}}$ of mesons in the ground state are significantly different. For example, for the ground state~($0^1S_0$) of the $u\bar{u}$ meson system, $\beta_{eff,\mathbf{r}}= 0.4760 $~GeV,~$\beta_{eff,\mathbf{p}} = 0.5405$~GeV; for the ground state~($0^1S_0$) of the $u\bar{s}$ meson system, $\beta_{eff,\mathbf{r}} = 0.4535 $~GeV,~~$\beta_{eff,\mathbf{p}} = 0.5024$~GeV.
From the above two sets of data, we can see that when the meson is in the ground state, the~$\beta_{eff}$ obtained in coordinate space and momentum space are significantly different, while in the highly excited states, $\beta_{eff,\mathbf{r}}$ and  $\beta_{eff,\mathbf{p}}$ are relatively close.

\vspace{20pt}
\hspace{60pt}
{{\it{case 2: variable $\beta$}}}
\vspace{10pt}
\par
Taking $\beta$ as a function of energy, the $\beta$ at the lowest ground state energy is obtained by the variational method. Then the $\beta_{eff}$ value in corresponding coordinate space and momentum space will produced by solving the Eq. (\ref{3.4}). As shown in Table \ref{tab3} and Table \ref{tab4},
the results obtained with the $\beta$  selected in this way  are consistent with   those obtained with $\beta$ = 0.4  GeV.
This further demonstrates the appropriateness of using $\beta=0.4 $ GeV as the input for the numerical calculation of $\beta_{eff}$ in all previous references. Similarly, we can still see from the Table \ref{tab3} and Table \ref{tab4} that $\beta_{eff,\mathbf{r}}$ differs greatly from $\beta_{eff,\mathbf{p}}$ in the ground state. The behavior of $\beta_{eff,\mathbf{r}}$ and $\beta_{eff,\mathbf{p}}$ are similar in the highly excited state.

\renewcommand{\arraystretch}{1.2}
\begin{table*}[htbp]
\caption{The $\beta_{eff}$ value of the $u\bar{u}$  quark system corresponding to case 2 (unit: GeV).}
\[\begin{array}{ccccc|ccccc}
\toprule[1pt] \toprule[1pt]
state  &  \beta & \beta_{eff,\mathbf{r}} &  \beta_{eff,\mathbf{r}}$ \cite{Close:2005se}$ &  \beta_{eff,\mathbf{p}} & state &  \beta & \beta_{eff,\mathbf{r}}  &  \beta_{eff,\mathbf{r}}  $ \cite{Close:2005se}$ &  \beta_{eff,\mathbf{p}}  \\
\hline
0^1S_0 & 0.7020      & 0.4783      & 0.47 & 0.5452 & 0^3S_1 & 0.4681      & 0.2837      & 0.28 & 0.2850 \\
1^1S_0 & 0.4876      & 0.2805      & 0.28 & 0.2718 & 1^3S_1 & 0.3121      & 0.2412      & 0.24 & 0.2391 \\
2^1S_0 & 0.4096      & 0.2413      & 0.24 & 0.2345 & 2^3S_1 & 0.3121      & 0.2221      & 0.23 & 0.2201 \\
3^1S_0 & 0.3121      & 0.2212      &      & 0.2154 & 3^3S_1 & 0.2926      & 0.2102      &      & 0.2084 \\
0^1P_1 & 0.4681      & 0.2715      & 0.27 & 0.2762 & 0^3P_J & 0.3901      & 0.2630      & 0.26 & 0.2644 \\
1^1P_1 & 0.3901      & 0.2387      &      & 0.2414 & 1^3P_J & 0.3121      & 0.2325      &      & 0.2319 \\
2^1P_1 & 0.3121      & 0.2214      &      & 0.2228 & 2^3P_J & 0.3121      & 0.2167      &      & 0.2156 \\
3^1P_1 & 0.3121      & 0.2102      &      & 0.2108 & 3^3P_J & 0.2731      & 0.2063      &      & 0.2050 \\
0^1D_2 & 0.3121      & 0.2468      & 0.25 & 0.2479 & 0^3D_J & 0.3121      & 0.2464      & 0.25 & 0.2474 \\
1^1D_2 & 0.3121      & 0.2250      &      & 0.2252 & 1^3D_J & 0.2926      & 0.2244      &      & 0.2245 \\
2^1D_2 & 0.3121      & 0.2121      &      & 0.2120 & 2^3D_J & 0.2731      & 0.2116      &      & 0.2111 \\
3^1D_2 & 0.2731      & 0.2031      &      & 0.2027 & 3^3D_J & 0.2537      & 0.2026      &      & 0.2019 \\
\bottomrule[1pt] \bottomrule[1pt]
\end{array}\]\label{tab3}
\end{table*}

\renewcommand{\arraystretch}{1.2}
\begin{table*}[htbp]
\caption{The $\beta_{eff}$ value of the $u\bar{s}$  quark system corresponding to case 2 (unit: GeV).}
\[\begin{array}{ccccc|ccccc}
\toprule[1pt] \toprule[1pt]
state  &  \beta & \beta_{eff,\mathbf{r}} &  \beta_{eff,\mathbf{r}} $ \cite{Close:2005se}$ &  \beta_{eff,\mathbf{p}} & state &  \beta & \beta_{eff,\mathbf{r}} &  \beta_{eff,\mathbf{r}} $ \cite{Close:2005se}$ &  \beta_{eff,\mathbf{p}} \\
\hline
0^1S_0 & 0.7020      & 0.4551      & 0.46 & 0.5057 & 0^3S_1 & 0.5655      & 0.3155      & 0.32 & 0.3180 \\
1^1S_0 & 0.5071      & 0.2946      & 0.29 & 0.2903 & 1^3S_1 & 0.4486      & 0.2638      & 0.26 & 0.2616 \\
2^1S_0 & 0.4291      & 0.2566      & 0.25 & 0.2518 & 2^3S_1 & 0.3121      & 0.2416      & 0.24 & 0.2395 \\
3^1S_0 & 0.3706      & 0.2371      &      & 0.2327 & 3^3S_1 & 0.3121      & 0.2280      &      & 0.2261 \\
0^1P_1 & 0.4681      & 0.2939      & 0.29 & 0.2985 & 0^3P_J & 0.3901      & 0.2859      & 0.29 & 0.2876 \\
1^1P_1 & 0.4291      & 0.2574      &      & 0.2596 & 1^3P_J & 0.3121      & 0.2518      &      & 0.2512 \\
2^1P_1 & 0.3121      & 0.2385      &      & 0.2393 & 2^3P_J & 0.3121      & 0.2344      &      & 0.2332 \\
3^1P_1 & 0.3121      & 0.2262      &      & 0.2263 & 3^3P_J & 0.2731      & 0.2229      &      & 0.2216 \\
0^1D_2 & 0.4096      & 0.2673      & 0.27 & 0.2685 & 0^3D_J & 0.3706      & 0.2668      & 0.27 & 0.2679 \\
1^1D_2 & 0.3121      & 0.2431      &      & 0.2434 & 1^3D_J & 0.2926      & 0.2426      &      & 0.2426 \\
2^1D_2 & 0.3121      & 0.2290      &      & 0.2288 & 2^3D_J & 0.2926      & 0.2285      &      & 0.2280 \\
3^1D_2 & 0.3121      & 0.2193      &      & 0.2188 & 3^3D_J & 0.2731      & 0.2187      &      & 0.2179 \\
\bottomrule[1pt] \bottomrule[1pt]
\end{array}\]\label{tab4}
\end{table*}

\begin{figure}[htp]
\begin{center}
	\centerline{\includegraphics[width=235pt]{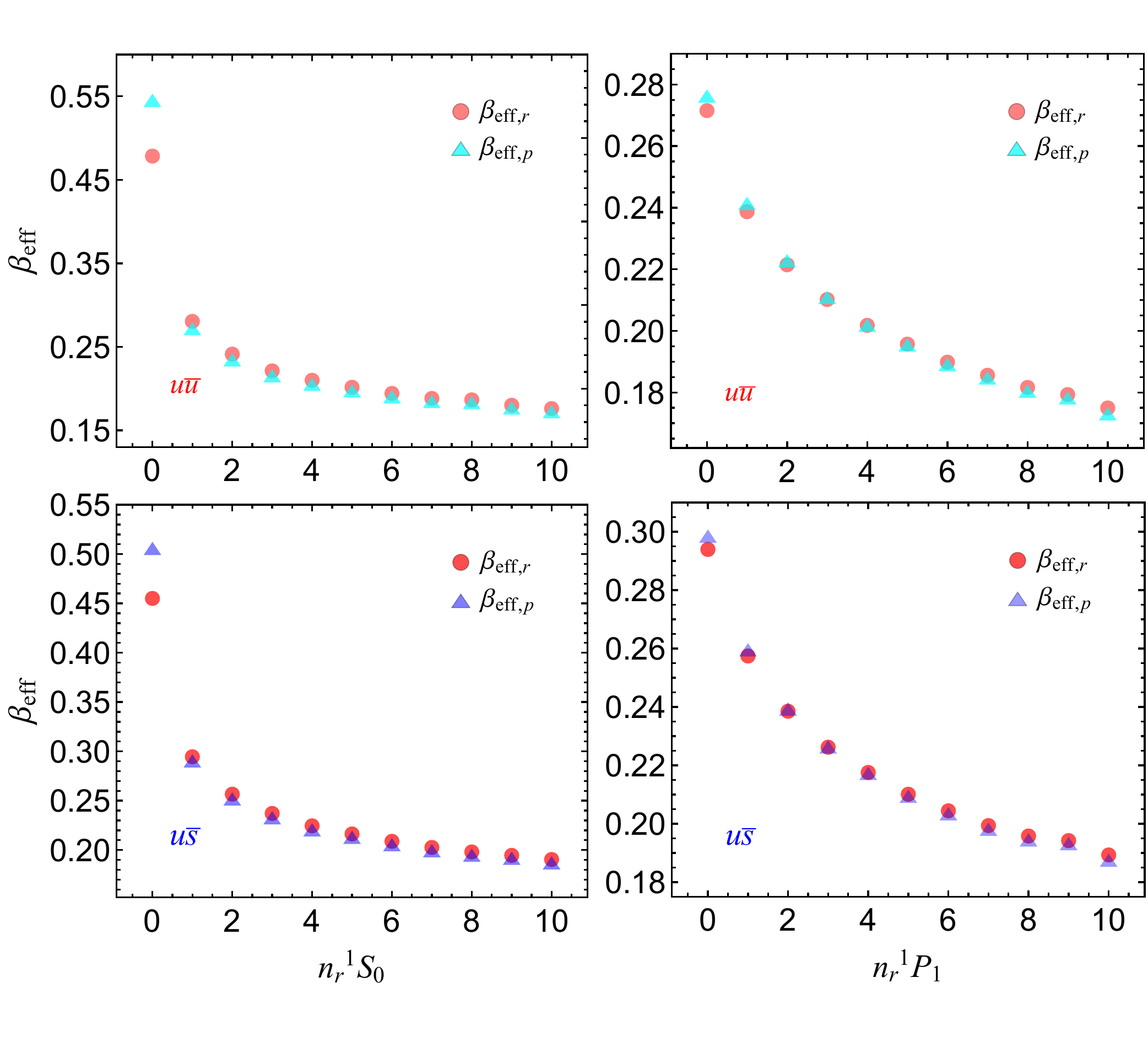}}
\caption{Graphs of $\beta_{eff}$ value of the $u\bar{u}$ and $u\bar{s}$  quark systems when $\beta$ is taken at lowest energy, where the abscissa represents the radial quantum number $n_r$, and the ordinate is the magnitude of $\beta_{eff,\mathbf{r}}$ and $\beta_{eff,\mathbf{p}}$, in units of GeV.}
\label{figuc}
\end{center}
\end{figure}

In addition,  we draw the curves of $\beta_{eff,\mathbf{r}}$ and $\beta_{eff,\mathbf{p}}$  for different radial quantum numbers ($n_r$) of the same orbital quantum number (L) in FIG. \ref{figuc}. From the figure, we can see that the values  of $\beta_{eff,\mathbf{r}}$ and $\beta_{eff,\mathbf{p}}$ coincide in the highly excited state. It is permissible to replace the behavior of $\beta_{eff,\mathbf{p}}$  with $\beta_{eff,\mathbf{r}}$  in the QPC model. However, the behavior of both in the ground state tells us that when using QPC to calculate the decay amplitude, it is necessary to be careful to select $\beta_{eff}$ as the calculation parameter if the effective SHO wave function is for the ground state.

We use $\beta_{eff,\textbf{r}}$ and  $\beta_{eff,\textbf{p}}$  as the parameter in the effective SHO wave function  of the QPC model to calculate several decay channels, which are in the ground states. The decay channels and quark mass obtained are from Ref. \cite{Ye:2012gu}.  The results are shown in  Table \ref{tabQPCcal}.

\renewcommand{\arraystretch}{1.2}
\begin{table}[htp]
\centering
 \footnotesize
	\caption{ Comparison of the decay widths calculated by QPC model with $\beta_{eff}$ from coordinate space and momentum space, and the results are obtained in the QPC model when $m_u=m_d=0.33$~GeV and $m_s=0.55$~GeV, the unit of width is MeV. }
\label{uu,0.4}
\[\begin{array}{ccccccccc}
\toprule[1pt]\toprule[1pt]
Decay~channel & Measured$\cite{ParticleDataGroup:2010dbb}$  & \beta_ {eff,\mathbf{r}} &\beta_ {eff,\mathbf{p}} \\
\hline
b_1(1235)\to \omega \pi     &  142 \pm 8       & 118.106    & 125.297 \\
\phi\to K^+K^-              &  2.08 \pm 0.02   & 1.908      & 2.010   \\
a_2(1320)\to \eta \pi       &  15.5 \pm 0.7    & 21.280     & 13.525  \\
a_2(1320)\to K\bar{K}       &  5.2 \pm 0.2     & 2.277      & 1.638   \\
\pi_2(1670)\to f_2(1270)\pi &  145.8 \pm 5.1   & 135.914    & 151.546 \\
\rho_3(1690)\to \pi\pi      &  38 \pm 2.4      & 32.801     & 15.528  \\
\rho_3(1690)\to \omega \pi  &  25.8 \pm 1.6    & 65.359     & 57.558  \\
\rho_3(1690)\to K\bar{K}    &  2.5 \pm 0.2     & 1.330      & 0.728   \\
K^*(892)\to K\pi            &  48.7 \pm 0.8    & 28.029     & 27.245  \\
K^*(1410)\to K\pi           &  15.3 \pm 1.4    & 11.307     & 27.623  \\
K^*_0(1430)\to K\pi         &  251 \pm 74      & 165.174    & 258.416 \\
K^*_2(1430)\to K\pi         &  54.4 \pm 2.5    & 64.803     & 49.188  \\
K^*_2(1430)\to K^*(892)\pi  &  26.9 \pm 1.2    & 31.526     & 29.538 \\
K^*_2(1430)\to K\rho        &  9.5 \pm 0.4     & 14.022     & 14.248  \\
K^*_2(1430)\to K\omega      &  3.16 \pm 0.15   & 4.292      & 4.361   \\
\hline
                            &                   & \chi^2=1913  & \chi^2=1914 \\
\hline
                            &                   & \gamma=8.663  & \gamma=9.871 \\
\bottomrule[1pt]\bottomrule[1pt]
\end{array}\]\label{tabQPCcal}
\end{table}

It can be seen from Table~\ref{tabQPCcal}  that $\beta_ {eff,\mathbf{r}}$ obtained from coordinate space and $\beta_ {eff,\mathbf{p}}$ from momentum space have great influence on the calculation results of decay width when a SHO wave function is taken as effective wave function in QPC model. Compared with Ref. \cite{Ye:2012gu}, the decay width calculated from coordinate space is close to the data of Ref. \cite{Ye:2012gu}.

What is more noteworthy in Table~\ref{tabQPCcal}  is that, we found that when $\beta_{eff}$ is taken from different spaces, it has a great influence on the vacuum generation rate $\gamma$  and decay width. Firstly, the $\gamma$ obtained by fitting in $\beta_ {eff,\mathbf{r}}$ is $12\%$ less than it by fitting $\beta_ {eff,\mathbf{p}}$.  The $\beta_ {eff,\mathbf{r}}$ and $\beta_ {eff,\mathbf{p}}$ of the highly excited states are roughly equal, but the $\gamma$ value obtained from different spaces are highly correlated with the decay width, which also makes the difference between the two reach about $20 \%$ in some decays containing the highly excited states. Secondly, the average relative error between the decay width obtained from the $\beta_ {eff,\mathbf{r}}$ and the $\beta_ {eff,\mathbf{p}}$  with the experimental value is calculated. The formula is as follows
\begin{equation}
\overline{RE}=\frac{1}{15}\sum_{i=1}^{15} \frac{|\Gamma_i(\beta_ {eff,\mathbf{r}})-\Gamma_i(\beta_ {eff,\mathbf{p}})|}{\Gamma_i(Exp)},
\end{equation}
and the $\overline{RE}=0.2459$. This means that the difference is about $25 \%$ using $\beta_ {eff,\mathbf{r}}$ and $\beta_ {eff,\mathbf{p}}$ to calculate the decay width, respectively.
This will have important implications for the use of QPC model.

\vspace{20pt}
\hspace{60pt}
{{\it{case 3: $\beta=\sqrt{\beta_{eff,\mathbf{r}}\beta_{eff,\mathbf{p}}}$ }}}
\vspace{10pt}
\par
We discuss equations (\ref{3.3}) and (\ref{3.4}) as follows. Since $\beta$ is theoretically arbitrary,  $\beta^2_{eff,\mathbf{r}}$ and   $\beta_{eff,\mathbf{p}}^2$  are independent of $\beta$. We know from (\ref{3.4}) that  $k_1-k_2$ is equal to  $\beta^2$ multiply some constant, and  $k_1+k_2$ is equal to  $1/\beta^2$ multiply some constant. The relation between $\beta_{eff,\mathbf{r}}$ and $\beta_{eff,\mathbf{p}}$ will be obtained:
\begin{equation}\label{3.11}
\begin{split}
\beta^2_{eff,\mathbf{r}}-\beta_{eff,\mathbf{p}}^2=\frac{\beta^2[(2n_r+L+\frac{3}{2})^2-(k_1^2-k_2^2)]}
{(2n_r+L+\frac{3}{2})(k_1-k_2)},
\end{split}
\end{equation}
or
\begin{equation}\label{3.12}
\begin{split}
\frac{\beta^2_{eff,\mathbf{r}}}{\beta_{eff,\mathbf{p}}^2}=\frac{(2n_r+L+\frac{3}{2})^2}{k_1^2-k_2^2}.
\end{split}
\end{equation}

The calculation is as follows from Eq. (\ref{3.4}) :
\begin{equation}\label{3.13}
\begin{split}
\begin{cases}
&\beta^{-2}_{eff,\mathbf{r}}(2n_r+L+\frac{3}{2})=\beta^{-2}{(k_1-k_2)},\\
&\beta_{eff,\mathbf{p}}^2(2n_r+L+\frac{3}{2})=\beta^2(k_1+k_2).
\end{cases}
\end{split}
\end{equation}
If let $\beta^2=\beta_{eff,\mathbf{p}}\beta_{eff,\mathbf{r}}$, the formula (\ref{3.13}) can be written as
\begin{equation}\label{3.14}
\begin{split}
\begin{cases}
&(2n_r+L+\frac{3}{2})\frac{\beta_{eff,\mathbf{p}}}{\beta_{eff,\mathbf{r}}}=k_1-k_2,\\
&(2n_r+L+\frac{3}{2})\frac{\beta_{eff,\mathbf{p}}}{\beta_{eff,\mathbf{r}}}=k_1+k_2.
\end{cases}
\end{split}
\end{equation}
We find that the left side of equation (\ref{3.14}) is equal, the right side is same only when $k_2= 0.$
i. e.
\begin{equation}\label{3.15}
\begin{cases}
&\beta=\sqrt{\beta_{eff,\mathbf{p}}\beta_{eff,\mathbf{r}}}.\\
&k_2=2\sum_{n=0}^{N-2}c_{n}^*c_{n+1}\sqrt{(n+1)(n+L+\frac{3}{2})}= 0\\
&\frac{\beta_{eff,\mathbf{p}}}{\beta_{eff,\mathbf{r}}}=\frac{k_1}{2n_r+L+\frac{3}{2}}=\frac{\sum_{n=0}^{N-1}c_{n}^* c_{n}(2n+L+\frac{3}{2})}{2n_r+L+\frac{3}{2}}.
\end{cases}
\end{equation}
From equation (\ref{3.15}), we also know that the average value of $n$ in the highly excited states ($\beta_{eff,\mathbf{p}}/\beta_{eff,\mathbf{r}}\approx 1$) is approximately equal to the radial quantum number $n_r$.

In the specific calculation, we can do a cycle, so that $\beta$ approaches the square root of $\sqrt{\beta_{eff,\mathbf{p}}\beta_{eff,\mathbf{r}}}$. We test the relation (\ref{3.15}) numerically. The cycles of~~$k_2<10^{-3}$,~ $\beta,~\beta_{eff,\mathbf{r}},\\~\beta_{eff,\mathbf{p}},~\sqrt{\beta_{eff,\mathbf{p}}\beta_{eff,\mathbf{r}}},~
k_2$ and $k_1$ are  listed in Table \ref{tab5} and Table \ref{tab6}. It can be seen from Table \ref{tab5} and Table \ref{tab6} that when $\beta=\sqrt{\beta_{eff,\mathbf{p}}\beta_{eff,\mathbf{r}}}$, the sizes of~~$\beta_{eff,\mathbf{r}}$ ~and ~$\beta_{eff,\mathbf{p}}$ in the ground states are quite different, the sizes of them in the excited states are  similar, which is consistent with the behavior of the previous two cases. The data in Table \ref{tab5} and Table \ref{tab6} also verifies the identity of $k_2 = 0$, and $k_1/2n_r+L+\frac{3}{2} \approx 1$, further verifies that the average value of $n$ is approximately equal to the radial quantum number $n_r$ at highly excited states.

\renewcommand{\arraystretch}{1.2}
\begin{table*}[htbp]
\centering
 \footnotesize
	\caption{Verification of the $k_2$  value in the $u\bar{u}$ quark system (S=0) and the $\beta_{eff}$ value when $k_2<10^{-3}$. }
\[\begin{array}{cccccccc}
\toprule[1pt]\toprule[1pt]
state & \beta & \beta_{eff,\mathbf{r}} & \beta_{eff,\mathbf{p}} & \sqrt{\beta_{eff,\mathbf{r}}\beta_{eff,\mathbf{p}}} & k_2 &
\frac{k_1}{2n_r+L+\frac{3}{2}} &  cycle~~index  \\
\hline
 0^1S_0     & 0.5099   & 0.4776 & 0.5443 & 0.5099 & 1.421\times 10^{-4} & 1.1395 & 2 \\
 1^1S_0     & 0.2721   & 0.2777 & 0.2666 & 0.2721 & 4.554\times 10^{-4} & 0.9600 & 2 \\
 2^1S_0     & 0.2339   & 0.2384 & 0.2294 & 0.2339 & 1.267\times 10^{-4} & 0.9621 & 2 \\
 3^1S_0     & 0.2158   & 0.2196 & 0.2121 & 0.2158 & 1.869\times 10^{-4} & 0.9662 & 2 \\
 4^1S_0     & 0.2044   & 0.2077 & 0.2013 & 0.2044 & 1.921\times 10^{-6} & 0.9693 & 2 \\
 5^1S_0     & 0.1961   & 0.1990 & 0.1933 & 0.1961 & 4.544\times 10^{-5} & 0.9712 & 3 \\
 6^1S_0     & 0.1895   & 0.1922 & 0.1868 & 0.1895 & 1.117\times 10^{-5} & 0.9718 & 3 \\
\hline
 0^1P_1     & 0.2733   & 0.2712 & 0.2754 & 0.2733 & 2.858\times 10^{-4} & 1.0157 & 2 \\
 1^1P_1     & 0.2375   & 0.2382 & 0.2401 & 0.2391 & 8.002\times 10^{-4} & 1.0080 & 2 \\
 2^1P_1     & 0.2213   & 0.2210 & 0.2216 & 0.2213 & 4.763\times 10^{-5} & 1.0028 & 3 \\
 3^1P_1     & 0.2095   & 0.2096 & 0.2094 & 0.2095 & 7.539\times 10^{-5} & 0.9991 & 3 \\
 4^1P_1     & 0.2008   & 0.2012 & 0.2005 & 0.2008 & 4.728\times 10^{-5} & 0.9963 & 3 \\
 5^1P_1     & 0.1940   & 0.1946 & 0.1934 & 0.1940 & 2.040\times 10^{-4} & 0.9939 & 3 \\
 6^1P_1     & 0.1882   & 0.1891 & 0.1874 & 0.1882 & 9.007\times 10^{-4} & 0.9913 & 3 \\
\hline
 0^1D_2     & 0.2474   & 0.2468 & 0.2479 & 0.2474 & 2.066\times 10^{-5} & 1.0042 & 2 \\
 1^1D_2     & 0.2251   & 0.2250 & 0.2252 & 0.2251 & 1.733\times 10^{-6} & 1.0010 & 2 \\
 2^1D_2     & 0.2120   & 0.2121 & 0.2118 & 0.2120 & 4.953\times 10^{-5} & 0.9988 & 2 \\
 3^1D_2     & 0.2028   & 0.2031 & 0.2026 & 0.2028 & 7.375\times 10^{-4} & 0.9974 & 2 \\
 4^1D_2     & 0.1959   & 0.1962 & 0.1955 & 0.1959 & 1.225\times 10^{-5} & 0.9964 & 3 \\
 5^1D_2     & 0.1903   & 0.1907 & 0.1898 & 0.1903 & 5.905\times 10^{-5} & 0.9954 & 3 \\
 6^1D_2     & 0.1855   & 0.1860 & 0.1849 & 0.1855 & 1.663\times 10^{-4} & 0.9943 & 3 \\
\hline
 0^1F_3     & 0.2353   & 0.2350 & 0.2356 & 0.2353 & 1.118\times10^{-6}  & 1.0028 & 2 \\
 1^1F_3     & 0.2180   & 0.2179 & 0.2181 & 0.2180 & 2.402\times 10^{-5} & 1.0010 & 2 \\
 2^1F_3     & 0.2069   & 0.2070 & 0.2069 & 0.2069 & 8.741\times 10^{-5} & 0.9992 & 2 \\
 3^1F_3     & 0.1990   & 0.1992 & 0.1987 & 0.1990 & 7.502\times 10^{-5} & 0.9979 & 2 \\
 4^1F_3     & 0.1927   & 0.1930 & 0.1924 & 0.1927 & 4.014\times 10^{-6} & 0.9969 & 2 \\
 5^1F_3     & 0.1876   & 0.1880 & 0.1873 & 0.1876 & 1.403\times 10^{-4} & 0.9961 & 2 \\
 6^1F_3     & 0.1833   & 0.1837 & 0.1828 & 0.1833 & 2.075\times 10^{-4} & 0.9951 & 3 \\
\hline
 0^1G_4     & 0.2267   & 0.2265 & 0.2270 & 0.2267 & 1.494\times 10^{-7} & 1.0021 & 2 \\
 1^1G_4     & 0.2126   & 0.2125 & 0.2127 & 0.2126 & 9.590\times10^{-8}  & 1.0013 & 2 \\
 2^1G_4     & 0.2030   & 0.2031 & 0.2030 & 0.2030 & 2.587\times 10^{-5} & 0.9998 & 2 \\
 3^1G_4     & 0.1959   & 0.1960 & 0.1958 & 0.1959 & 1.335\times 10^{-4} & 0.9986 & 2 \\
 4^1G_4     & 0.1902   & 0.1905 & 0.1900 & 0.1902 & 5.887\times 10^{-4} & 0.9977 & 2 \\
 5^1G_4     & 0.1855   & 0.1858 & 0.1853 & 0.1855 & 6.115\times10^{-6}  & 0.9969 & 3 \\
 6^1G_4     & 0.1815   & 0.1819 & 0.1811 & 0.1815 & 1.900\times 10^{-4} & 0.9960 & 3 \\
\hline
 0^1H_5     & 0.2200   & 0.2198 & 0.2202 & 0.2200 & 1.095\times 10^{-7} & 1.0018 & 2 \\
 1^1H_5     & 0.2081   & 0.2080 & 0.2083 & 0.2081 & 1.654\times 10^{-6} & 1.0014 & 2 \\
 2^1H_5     & 0.1997   & 0.1997 & 0.1997 & 0.1997 & 1.216\times 10^{-6} & 1.0002 & 2 \\
 3^1H_5     & 0.1932   & 0.1933 & 0.1932 & 0.1932 & 2.163\times 10^{-4} & 0.9992 & 2 \\
 4^1H_5     & 0.1880   & 0.1882 & 0.1879 & 0.1880 & 7.613\times 10^{-5} & 0.9983 & 2 \\
 5^1H_5     & 0.1837   & 0.1839 & 0.1834 & 0.1837 & 6.980\times 10^{-6} & 0.9976 & 3 \\
 6^1H_5     & 0.1799   & 0.1802 & 0.1796 & 0.1799 & 2.710\times 10^{-4} & 0.9967 & 3 \\
\bottomrule[1pt] \bottomrule[1pt]
\end{array}\]\label{tab5}
\end{table*}

\renewcommand{\arraystretch}{1.2}
\begin{table*}[htbp]
\centering
 \footnotesize
	\caption{Verification of the $k_2$  value in the $u\bar{u}$ quark system (S=1) and the $\beta_{eff}$ value when $k_2<10^{-3}$. }
\[\begin{array}{cccccccc}
\toprule[1pt]\toprule[1pt]
state & \beta & \beta_{eff,\mathbf{r}} & \beta_{eff,\mathbf{p}} & \sqrt{\beta_{eff,\mathbf{r}}\beta_{eff,\mathbf{p}}} & k_2  &\frac{k_1}{2n_r+L+\frac{3}{2}}  & cycle~~index  \\
\hline
 0^3S_1     & 0.2843   & 0.2837 & 0.2850 & 0.2843 & 1.053\times 10^{-5} & 1.0047 & 2 \\
 1^3S_1     & 0.2401   & 0.2411 & 0.2391 & 0.2401 & 2.090\times 10^{-5} & 0.9912 & 2 \\
 2^3S_1     & 0.2211   & 0.2221 & 0.2201 & 0.2211 & 3.843\times 10^{-6} & 0.9912 & 2 \\
 3^3S_1     & 0.2093   & 0.2102 & 0.2084 & 0.2093 & 1.931\times 10^{-5} & 0.9917 & 2 \\
 4^3S_1     & 0.2008   & 0.2016 & 0.2000 & 0.2008 & 4.245\times 10^{-5} & 0.9921 & 2 \\
 5^3S_1     & 0.1943   & 0.1951 & 0.1936 & 0.1943 & 4.456\times 10^{-6} & 0.9923 & 3 \\
 6^3S_1     & 0.1890   & 0.1897 & 0.1882 & 0.1890 & 3.031\times 10^{-5} & 0.9922 & 3 \\
\hline
 0^3P_J     & 0.2637   & 0.2630 & 0.2644 & 0.2637 & 1.385\times 10^{-6} & 1.0053 & 2 \\
 1^3P_J     & 0.2322   & 0.2325 & 0.2319 & 0.2322 & 6.850\times 10^{-7} & 0.9975 & 2 \\
 2^3P_J     & 0.2162   & 0.2167 & 0.2156 & 0.2162 & 1.372\times 10^{-5} & 0.9948 & 2 \\
 3^3P_J     & 0.2057   & 0.2063 & 0.2050 & 0.2057 & 2.765\times 10^{-4} & 0.9938 & 2 \\
 4^3P_J     & 0.1980   & 0.1986 & 0.1973 & 0.1980 & 3.343\times 10^{-4} & 0.9932 & 2 \\
 5^3P_J     & 0.1919   & 0.1926 & 0.1912 & 0.1919 & 2.344\times 10^{-5} & 0.9929 & 3 \\
 6^3P_J     & 0.1869   & 0.1876 & 0.1862 & 0.1868 & 2.647\times 10^{-5} & 0.9923 & 3 \\
\hline
 0^3D_J     & 0.2469   & 0.2464 & 0.2474 & 0.2469 & 4.033\times 10^{-8} & 1.0038 & 2 \\
 1^3D_J     & 0.2245   & 0.2244 & 0.2244 & 0.2245 & 8.268\times 10^{-7} & 1.0002 & 2 \\
 2^3D_J     & 0.2113   & 0.2116 & 0.2111 & 0.2113 & 1.469\times 10^{-5} & 0.9978 & 2 \\
 3^3D_J     & 0.2022   & 0.2026 & 0.2019 & 0.2022 & 1.606\times 10^{-4} & 0.9964 & 2 \\
 4^3D_J     & 0.1953   & 0.1957 & 0.1948 & 0.1953 & 9.824\times 10^{-4} & 0.9954 & 2 \\
 5^3D_J     & 0.1897   & 0.1902 & 0.1892 & 0.1897 & 1.521\times 10^{-5} & 0.9946 & 3 \\
 6^3D_J     & 0.1850   & 0.1856 & 0.1845 & 0.1850 & 1.251\times 10^{-4} & 0.9938 & 3 \\
\hline
 0^3F_J     & 0.2353   & 0.2349 & 0.2356 & 0.2353 & 2.065\times10^{-7}  & 1.0027 & 2 \\
 1^3F_J     & 0.2179   & 0.2178 & 0.2180 & 0.2179 & 1.071\times 10^{-5} & 1.0010 & 2 \\
 2^3F_J     & 0.2069   & 0.2070 & 0.2068 & 0.2069 & 4.954\times 10^{-5} & 0.9991 & 2 \\
 3^3F_J     & 0.1989   & 0.1991 & 0.1987 & 0.1989 & 5.762\times 10^{-5} & 0.9978 & 2 \\
 4^3F_J     & 0.1927   & 0.1930 & 0.1923 & 0.1927 & 2.722\times 10^{-6} & 0.9968 & 3 \\
 5^3F_J     & 0.1876   & 0.1880 & 0.1873 & 0.1876 & 5.322\times 10^{-5} & 0.9959 & 2 \\
 6^3F_J     & 0.1832   & 0.1837 & 0.1828 & 0.1832 & 1.799\times 10^{-5} & 0.9950 & 3 \\
\hline
 0^3G_J     & 0.2267   & 0.2265 & 0.2270 & 0.2268 & 1.029\times 10^{-7} & 1.0021 & 2 \\
 1^3G_J     & 0.2126   & 0.2125 & 0.2127 & 0.2126 & 7.733\times10^{-8}  & 1.0013 & 2 \\
 2^3G_J     & 0.2030   & 0.2031 & 0.2030 & 0.2030 & 2.359\times 10^{-5} & 0.9998 & 2 \\
 3^3G_J     & 0.1959   & 0.1960 & 0.1958 & 0.1959 & 1.255\times 10^{-4} & 0.9986 & 2 \\
 4^3G_J     & 0.1902   & 0.1904 & 0.1900 & 0.1902 & 5.693\times 10^{-4} & 0.9977 & 2 \\
 5^3G_J     & 0.1855   & 0.1858 & 0.1852 & 0.1855 & 5.898\times10^{-6}  & 0.9968 & 3 \\
 6^3G_J     & 0.1815   & 0.1819 & 0.1811 & 0.1815 & 1.850\times 10^{-4} & 0.9960 & 3 \\
\hline
 0^3H_J     & 0.2200   & 0.2198 & 0.2202 & 0.2200 & 1.016\times 10^{-7} & 1.0018 & 2 \\
 1^3H_J     & 0.2081   & 0.2080 & 0.2083 & 0.2081 & 1.601\times 10^{-6} & 1.0014 & 2 \\
 2^3H_J     & 0.1997   & 0.1997 & 0.1997 & 0.1997 & 1.199\times 10^{-6} & 1.0003 & 2 \\
 3^3H_J     & 0.1932   & 0.1933 & 0.1932 & 0.1932 & 2.149\times 10^{-4} & 0.9992 & 2 \\
 4^3H_J     & 0.1880   & 0.1882 & 0.1879 & 0.1880 & 7.570\times 10^{-5} & 0.9983 & 2 \\
 5^3H_J     & 0.1837   & 0.1839 & 0.1834 & 0.1837 & 6.946\times 10^{-6} & 0.9976 & 3 \\
 6^3H_J     & 0.1799   & 0.1802 & 0.1796 & 0.1799 & 2.702\times 10^{-4} & 0.9967 & 3 \\
\bottomrule[1pt] \bottomrule[1pt]
\end{array}\]\label{tab6}
\end{table*}

\section{Conclusion}\label{sec4}

We derived the analytical expressions of $\beta_{eff,\mathbf{r}}$ and $\beta_{eff,\mathbf{p}}$ in the coordinate space and momentum space respectively, and verify the correctness of the analytical expressions by using the hydrogen atom wave function as the exact wave function. By applying these expressions to $u\bar{u}$ and $u\bar{s}$ quark systems and calculating their $\beta_{eff,\mathbf{r}}$ and $\beta_{eff,\mathbf{p}}$ in different states, we find that our results in coordinate space are consistent with the Reference  \cite{Close:2005se}.

Moreover, we compare in detail the behavior of $\beta_{eff,\mathbf{r}}$ and $\beta_{eff,\mathbf{p}}$ when $\beta$ in the exact wave function takes different values under Cornell potential.
We find that there is a significant difference between $\beta_{eff,\mathbf{r}}$ and $\beta_{eff,\mathbf{p}}$ in the ground states, while they are close in the highly excited states. It is reasonable to replace the value of $\beta_{eff,\mathbf{r}}$ with $\beta_{eff,\mathbf{p}}$  in the highly excited states, and vice versa, not including the ground state. The reason for this behavior is caused by Cornell potential in Hamiltonian. In the ground state, the Coulomb potential plays a leading role, while in the single Coulomb potential, when $L = 0$, the ratio of $\beta_ {eff,\mathbf{r}}/\beta_{eff,\mathbf{p}}$ is not equal to 1. In the highly excited state, the linear potential plays a leading role, so it is consistent with the value of single linear potential $\beta_ {eff,\mathbf{r}}/\beta_{eff,\mathbf{p}}$, that is, it is approximately 1. A specific example is when calculating the $\beta_ {eff,\mathbf{r}}/\beta_{eff,\mathbf{p}}$ under the Coulomb potential that the hydrogen atom wave function needs to satisfy as an exact wave function in Sec. \ref{sec2.3}, there is an obvious difference in the behavior of $\beta_ {eff,\mathbf{r}}/\beta_{eff,\mathbf{p}}$ in the ground state when $n_r$ and $L$ tend to infinity.

In addition, we discuss the behavior of $\beta_{eff,\mathbf{r}}$ and $\beta_{eff,\mathbf{p}}$  in $\beta=\sqrt{~\beta_{eff,\mathbf{r}}\beta_{eff,\mathbf{p}}}$  as input. We got an identity $k_2=0$ and $\bar{n}=n_r$ (for highly exited states), which tells us that the mean value of $n$ at $\beta=\sqrt{\beta_ {eff,\mathbf{r}}\beta_{eff,\mathbf{p}}}$ is approximately equal to the radial quantum number $n_r$ in the highly excited states. It is worth noting that this conclusion only holds for Cornell potential and does not hold for other potential.

It is concluded that when a SHO wave function is used as the effective wave function, the $\beta_{eff}$ value of coordinate space and momentum space can be obtained according to the analytical expression in SHO bases. We hope that our derivation will help to improve  the theoretical calculation results when using a SHO wave function as the effective wave function.

\section{ACKNOWLEDGMENTS}

This work is supported  by the National Natural Science Foundation of China under Grants No. 11965016,  National Natural Science Foundation of China, under Grants No. 12047501, the projects funded by Science and Technology Department of Qinghai Province (No. 2020-ZJ-728,No. 2018-ZJ-971Q).

%

\begin{widetext}

\end{widetext}

\bibliographystyle{apsrev4-1}
\bibliography{ref}
\end{document}